\documentclass[10pt,journal,compsoc]{IEEEtran}

\usepackage[nocompress]{cite}
\usepackage{url}
\usepackage{booktabs}
\usepackage{amssymb}
\usepackage{multirow}
\usepackage{enumitem}
\usepackage{adjustbox}
\usepackage{balance}
\usepackage{xcolor}
\usepackage[colorlinks,bookmarksopen,bookmarksnumbered,citecolor=red,urlcolor=red]{hyperref}
\usepackage{fontawesome5}
\usepackage{tabularray}

\usepackage{subfig}

\usepackage{epigraph}

\graphicspath{{figures}{authors}{.}}

\definecolor{cog}{HTML}{4A77B4}
\definecolor{ctx}{HTML}{5B9F5B}
\definecolor{int}{HTML}{D97B3A}
\definecolor{plat}{HTML}{888888}
\definecolor{vis}{HTML}{CC6666}
\definecolor{collab}{HTML}{8866AA}
\definecolor{eval}{HTML}{449999}

\newcommand{\mypara}[1]{\par\smallskip\noindent\textbf{#1}\enspace\ignorespaces}

\begin{document}


\title{Somewhere Over the Desktop: A Research Agenda for Ubiquitous Analytics}

\author{Niklas Elmqvist,~\IEEEmembership{Fellow,~IEEE,} Panagiotis D.\ Ritsos,~\IEEEmembership{Senior Member,~IEEE,} and Peter W.\ S.\ Butcher%
\IEEEcompsocitemizethanks{\IEEEcompsocthanksitem Niklas Elmqvist is with Aarhus University, Aarhus, Denmark.\protect\\
E-mail: elm@cs.au.dk
\IEEEcompsocthanksitem Panagiotis D.\ Ritsos and Peter W.\ S.\ Butcher are with Bangor University, Bangor, Wales, United Kingdom.\protect\\
E-mail: \{p.ritsos, p.butcher\}@bangor.ac.uk
}}

\IEEEtitleabstractindextext{%
\begin{abstract}
    Spatial computing, generative AI, and open web standards are converging. 
Three spatial operating systems---Android XR, Meta Horizon OS, and Apple visionOS---now ship with platform-level scene understanding.
Wearable displays span the range from full headsets to slim smartglasses. 
Agentic AI operates on the same spatial substrates as the human user.
This convergence enables new opportunities for \textit{ubiquitous analytics} (UA): the use of many, physically distributed, networked devices to support data sensemaking anytime and anywhere.
But proprietary platforms are settling design conventions that will calcify without evidence-based alternatives.
UA has now matured to the point where its intellectual history can be read as a structured genealogy of foundations, contributions, and lineages.
We trace this genealogy and organize it into clusters spanning cognition, context, interaction, platforms, visualization, collaboration, and evaluation.
Finally, we cross these clusters with each other, yielding a total of 42 future research challenges.
\end{abstract}

}

\maketitle

\epigraph{\itshape``Convergence,'' Tool said. ``Power ever draws other power.''}%
         {--- Steven Erikson, \emph{Gardens of the Moon} (1999)}

\section{Introduction}

We're seeing a convergence of great forces in spatial computing in the third decade of the new millenium. 
There are now three separate spatial OSes vying for power and market share: Apple's visionOS, Meta's HorizonOS, and Google's AndroidXR.
While full head-mounted display solutions such as the Apple Vision Pro are still struggling with acceptance, the Samsung Galaxy XR headset is already out, and slim form-factor XR glasses such as the Meta Ray-Bans, Xreal, and Rokid are shipping.
And the next big thing---the integration between AI and XR to enable an entirely new form of spatial computing---is being showcased in new products.
These developments are all heralding transformative potential in \textit{ubiquitous analytics}~\cite{Elmqvist2013}: the use of multiple, networked, and always-on digital devices that harnesses ubiquitous computing~\cite{Weiser1991} for the purpose of enabling data analytics anytime, anywhere~\cite{Elmqvist2023}.
But this great convergence also comes with great risk: in the struggle for market share, industrial giants are making decisions that threaten to calcify the future of APIs, platforms, hardware, use cases, and tasks for spatial computing for the foreseeable future. 

In this paper, we present an updated research agenda for ubiquitous analytics: seven positions on where the field should go, each unpacked into a handful of concrete research challenges.
The positions range from the analyst's mind outward to the infrastructure beneath their feet:
(1) that UA is grounded in distributed cognition, reflecting how analysts already think with the artifacts and devices around them~\cite{Hutchins1995, DBLP:journals/tochi/HollanHK00};
(2) that the physical environment, now legible to AI, becomes a computational substrate for analysis;
(3) that interaction must survive contact with the real world, where every input channel meets noise, motion, and constraint;
(4) that open web standards~\cite{DBLP:journals/cga/ButcherBSMER24} are the right foundation;
(5) that spatial visualization design is still in its infancy~\cite{Bressa2022};
(6) that analytical work is social and asymmetric; and
(7) that evaluation must leave the laboratory.
We derive the challenges by tracing the genealogy of prior art: where each line of work came from, what it inherited, and what it left unsolved.
Each challenge is anchored to a gap that the prior work opened but did not close.

The closest predecessor to this agenda is the set of immersive analytics grand challenges from Ens et al.~\cite{Ens2021}, drawn from an ACM CHI workshop in 2020.
Much has changed since.
The scope has widened: immersion is one corner of ubiquitous analytics, which also covers mobile, situated, and cross-device work that needs no headset at all~\cite{Elmqvist2013, Roberts2014}.
And technology has shifted: where spatial computing was a single-vendor bet in 2020, three spatial operating systems now ship at once, new generative AI models read rooms and objects as a platform service, and open web standards have begun to close the gap with proprietary engines~\cite{DBLP:journals/cga/ButcherBSMER24}.
Accordingly, our approach uses the lineage of research to identify the above seven themes that run through the field, then asks where they lead.
We argue that the future of ubiquitous analytics lies not in any single theme but in their \textit{convergence}: the questions that matter most sit where two or more themes meet.
We therefore study the themes not in isolation but pairwise---each against every other---and read the open challenges out of those intersections.

The contributions of this paper are the following:
(1) seven themes that run through ubiquitous analytics research---Cognition and the Analyst, With the World as My Context, Interaction Beyond the Prototype, Platforms and the Open Standards Race, Visualization in Space, Anytime Anywhere Together, and Evaluation in the Wild---each identified by tracing the lineage of prior work;
(2) a position on each theme about where it should go; and
(3) 42 research challenges, one for each ordered pair of themes, derived from the convergence of the two themes that produce it.

The remainder of this paper is structured as follows.
We first situate ubiquitous analytics and its prior research agendas. 
The next section introduces the seven themes, tracing the lineage of each and stating the position we take on it.
What follows is the core of the paper: the 42 challenges, organized as a theme-by-theme matrix, each challenge read out of the convergence of a pair of themes. 
We then step back to the cross-cutting questions: how theory meets evidence, how this relates to past grand challenges, and where AI fits in the big picture (spoiler: it's a big part of future research). 

All materials associated with this article can be found on OSF: \url{https://osf.io/32u7y}.
\section{Background}
\label{sec:background}

This agenda is based on three main topics:
ubiquitous analytics, cognitive frameworks, and earlier agendas.

\subsection{The Rise of Ubiquitous Analytics}

Weiser's vision of ubiquitous computing~\cite{Weiser1991} put computation into the world rather than on the desk:
not one machine commanding our attention but many, woven into the environment until they receded into the background.
Three decades on, this prediction has mostly come true.
We carry computation everywhere, but through one personal device rather than Weiser's scattered tabs and pads---\textit{quality computing} rather than \textit{quantity}, in Harrison et al.'s~\cite{DBLP:journals/ieeemm/HarrisonWD10} phrasing.
Elmqvist and Irani~\cite{Elmqvist2013} coined the term \textit{ubiquitous analytics} for sensemaking conducted anytime and anywhere across a collection of networked devices, and the later synthesis~\cite{Elmqvist2023} framed it as the analytical descendant of Weiser's program: data collected from everywhere and accessed from anywhere.
Roberts et al.~\cite{Roberts2014} gathered the strands of visualization beyond the desktop---mobile, large displays, tangible surfaces, and mixed reality---and called the move the next big thing, putting immersive and spatial displays on the agenda alongside the screens we already carried.

Two strands grew from this initial effort, each emphasizing a different setting.
\textit{Immersive analytics}, named by Chandler et al.~\cite{Chandler2015} and developed in the Marriott et al.~\cite{Marriott2018} book, takes the headset and the third spatial dimension as its starting point.
\textit{Situated analytics}~\cite{elsayed16situateddef} ties data to the physical referents it describes, and Willett et al.'s embedded representations~\cite{willett17embedded} push this until visualization and referent share the same space.
Bressa et al.~\cite{Bressa2022} survey how ``situatedness'' has been used in visualization.
Mobile data visualization, the subject of Lee et al.'s~\cite{Lee2022} book, needs no headset at all, and grew on its own branch as a result.

We treat these as regions of a single space rather than rival fields, because analysts move between them inside one task.
A field scientist consults a phone on the walk in, spreads dashboards across office monitors at the desk, and puts on a headset to inspect a model in 3D, all in service of one question.
An agenda scoped to immersion alone would have nothing to say about two-thirds of that day.

\subsection{Cognitive Foundation}

Visualization is fundamentally grounded in \textit{external cognition}~\cite{Scaife1996}; the use of external representations to offload memory, constrain computation, and re-represent data.
Analogously, from the very beginning, ubiquitous analytics has been defined on a foundation of post-cognitive frameworks~\cite{Elmqvist2023}, \textit{distributed cognition} (DCog) in particular. 
DCog was proposed by Hutchins~\cite{Hutchins1995} from his study of ship navigation and introduced to HCI by Hollan et al.~\cite{DBLP:journals/tochi/HollanHK00}.
The framework holds that human cognition is not restricted to our brains; it encompasses people, artifacts, and the environment, and the unit of analysis is the whole system.
The navigator computing a fix does not hold the computation in his head; it lives in the charts, the instruments, the division of labor, and the propagation of information between them.
Liu et al.~\cite{Liu2008} brought this framework to visualization, showing how external representations carry cognitive work that would otherwise fall on memory and attention.

DCog earns its place because it provides a reasoning framework for how sensemaking can be scaffolded by distributing the data analysis across instrumented components and collaborators in the analyst's surroundings, not merely the contents of a single computer screen.
When an analyst offloads a partial result onto a wall display, walks away, and returns an hour later, the wall is holding part of the thought.
When two analysts share a tabletop, the surface coordinates a computation neither performs alone.
Our prior UA work sharpen this lens into specific sensemaking \textit{instruments}~\cite{Beaudouin-Lafon2000, Beaudouin-Lafon2004}: a model of sensemaking as channels carried over substrates~\cite{Mackay2025substrates}, and a model of analytical reasoning as the construction of schemas on whatever substrates a setting affords, both building on Norman's gulfs~\cite{Norman1986} and their visualization-specific extension~\cite{Lam2008}.

\subsection{Grand Challenges in HCI and Visualization}

Charting a field's open problems is a recurring exercise in computing, and the grand-challenges paper is its standard form.
Shneiderman et al.~\cite{DBLP:journals/interactions/ShneidermanPCJE16} set out challenges for HCI at large;
Thomas and Cook's \emph{Illuminating the Path}~\cite{Thomas2005} defined the visual analytics agenda and the field it named;
and Johnson~\cite{DBLP:journals/cga/Johnson04} did the same for scientific visualization.

On immersive analytics,  Ens et al.'s survey drawn from a CHI 2020 workshop and refined by two dozen experts, present seventeen grand challenges in immersive analytics~\cite{Ens2021} across four themes: spatially situated visualization, interaction, collaborative analytics, and scenarios and evaluation.
We share the goal and depart on three counts, and the departures motivate the rest of the paper.
First, scope: Ens et al.\ address immersive analytics, one region of the space; we address UA, including the mobile and cross-device work that needs no headset.
Second, method: their challenges emerged from workshop consensus and are grouped by topic, while ours are read out of a stated position and each falls out of a convergence between two themes rather than standing alone.
Third, timing: their survey captures the field circa 2020, when spatial computing was effectively a single-vendor bet.
The convergence we described in the introduction had not happened yet.
Much of what was speculative then ships now, and the window that opens with it is the reason to define a new agenda.

\begin{figure*}[htb]
    \centering
    \includegraphics[width=\linewidth]{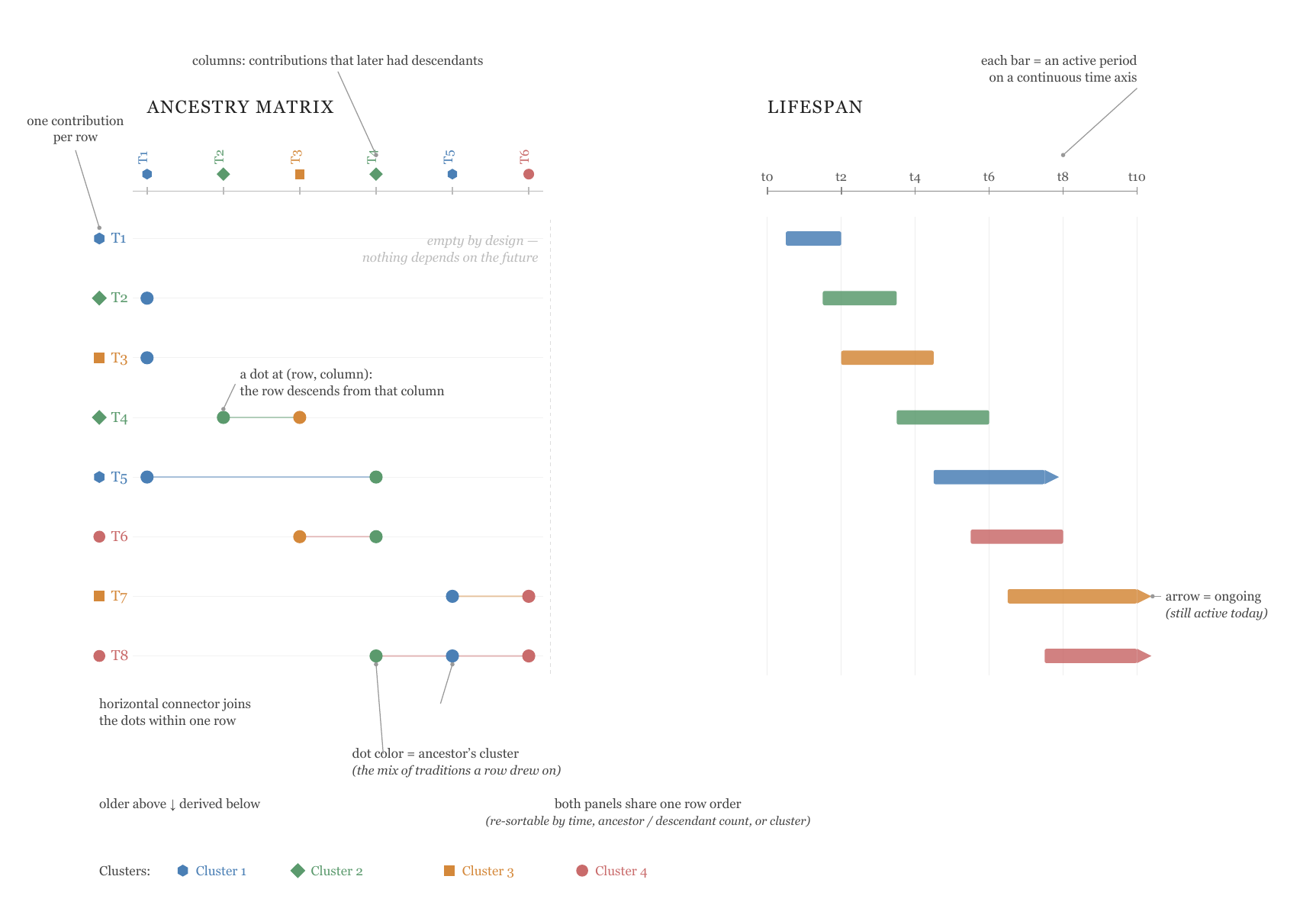}
    \caption{\textbf{Lineage plots.}
    Left: an ancestry matrix where each row is a contribution, columns are contributions with descendants, and a dot at (row, column) means the row descends from that column; a connector joins a row's ancestors, and each dot is colored by its ancestor's cluster.
    Right: a timeline panel sharing the row order, with active spans on a continuous time axis and arrows for ongoing work.
    }
    \label{fig:lineage-plots}
\end{figure*}

\section{Method}
\label{sec:method}

Our approach has three parts: bounding the corpus through expert-curated forward snowballing~\cite{Wohlin2014}, visualizing the result as a lineage plot of seven high-level themes, and generating research challenges by crossing themes pairwise.
Our procedure involved selecting a seed set of foundational contributions that the field routinely cites; tracing forward through the work that built on them; recording a descent link where a later contribution builds on a concept, technique, or framework of an earlier one (mere citation does not qualify); and coding each contribution into one of seven thematic clusters by author consensus.
We stopped tracing when new candidates no longer added inter-theme links or altered the cluster structure.

\subsection{Scope and Selection}
\label{sec:scope}

This paper traces the intellectual genealogy of ubiquitous analytics: what ideas the field inherited, where they branched, and where branches have yet to cross.
Rather than a systematic literature search, the goal of this paper is to identify seminal papers in the field and trace their lineage.
In other words, this is a genealogical problem based on domain knowledge that keyword search cannot supply.

Our process is a form of forward snowballing~\cite{Wohlin2014}, curated by expert judgment rather than exhaustive citation chasing, and our lineage plots are hand-curated analogues of main path analysis in citation networks~\cite{Hummon1989}.
The resulting corpus comprises 89 contributions connected by 154 links, spanning the years 1986 to 2026 across literature in visualization, HCI, and ubiquitous/immersive/situated analytics.

As a coverage check, the main ubiquitous analytics corpus (i.e., in the 2013--2026 period) includes every contribution discussed in at least two of five prior syntheses of the area~\cite{Ens2021, Marriott2018, Bressa2022, Lee2022, Shin2024}.

We bounded the corpus by intellectual descent rather than by coverage.
Starting from the foundational contributions the field routinely cites---Hutchins on distributed cognition~\cite{Hutchins1995}, Bertin on visual variables~\cite{Bertin1967}, Norman on the gulfs~\cite{Norman1986}, Weiser on ubiquitous computing~\cite{Weiser1991}---we traced forward through the work that built on them, using our combined knowledge of the visualization, HCI, immersive analytics, and situated analytics literatures.

We do not claim exhaustive coverage; a different author team with different expertise may have drawn some links differently and included contributions we omitted.
Expert judgment is the norm for this genre---the seventeen challenges of Ens et al.~\cite{Ens2021} emerged from workshop consensus among two dozen experts---but that judgment is rarely auditable after the fact.
To support this, we provide the node catalog, descent links, and cluster assignments on OSF,\footnote{Ubiquitous analytics lineage data (public): \url{https://osf.io/32u7y}} including the script that generates the plots in this paper.

We coded each contribution into one of seven thematic clusters: cognition, context, interaction, platforms, visualization, collaboration, and evaluation.
The clusters were not derived through formal open coding; we proposed candidate clusters reflecting the community's own disciplinary structure, assigned contributions to them independently, and resolved disagreements by discussion until the structure was stable.
We revisit these seven clusters in Section~\ref{sec:identification} as the themes the rest of the paper takes positions on.

\subsection{Lineage Plots}
\label{sec:lineage-plots}

To trace descent at the scale of a whole field, we propose a \textit{lineage plot} (Figure~\ref{fig:lineage-plots}) consisting of two panels sharing a single row order; one scientific contribution per row.
The left panel is an \emph{ancestry matrix}: rows are all contributions, columns are those contributions that went on to have descendants, and a dot at $(\textit{row}, \textit{column})$ means the row descends from the column.
A horizontal connector joins the dots within a row, so a single left-to-right scan gives a contribution's complete ancestry.
The approach draws inspiration from UpSet~\cite{DBLP:journals/tvcg/LexGSVP14} and GeneaQuilts~\cite{DBLP:journals/tvcg/BezerianosDFBW10}, respectively.
UpSet shows set intersections as a matrix of dots with a connector line per row and a bar chart of intersection sizes alongside; its central virtue is that each row is self contained.
We keep the dots and the connectors, and combine them with the second panel:
a Gantt-style \emph{lifespan} chart, where each contribution's active period is a bar on a continuous time axis and ongoing contributions end in a right-pointing arrow.
Similar to GeneaQuilts~\cite{DBLP:journals/tvcg/BezerianosDFBW10}, the ancestry matrix is genealogical rather than set-theoretic.

The plot is lower-triangular by default---older contributions at the top, derived contributions below, every dot to the left of its row.
The upper-right triangle stays empty on purpose; nothing depends on the future.

Color does the thematic work.
Each contribution belongs to one of seven \emph{clusters}, and three things carry that cluster: the row's label, its lifespan bar, and---this is the useful part---the dots in the columns.
A dot is colored by its \emph{ancestor's} cluster, so reading across a row shows the chromatic mix of traditions a contribution drew on.
A row with cognition-blue, visualization-green, and platform-orange dots inherited from three families at once.

We have released our D3 implementation of lineage plots on GitHub, where also the full ubiquitous analytics lineage can be explored (interactive versions of the plots in this paper): \url{https://nickelm.github.io/lineage-plot/}

\subsection{Future Challenges from Convergence}
\label{sec:pairwise}

The seven themes identified by the lineage analysis are the input to the challenge generation (Section~\ref{sec:challenges}).
Rather than enumerate open problems within each theme, we cross every theme with every other.
\textbf{The rationale is that future research will center around the convergence of research themes rather than on progress within any single one.}

With seven themes and no self-intersection, the method yields $7 \times 6 = 42$ challenges (Table~\ref{tab:challenge-matrix}).
These challenges are described in Section~\ref{sec:challenges}.
We do not claim that they are exhaustive; we return to what is missing in Section~\ref{sec:discussion}. 
\begin{figure*}[htb]
    \centering
    \includegraphics[width=\linewidth]{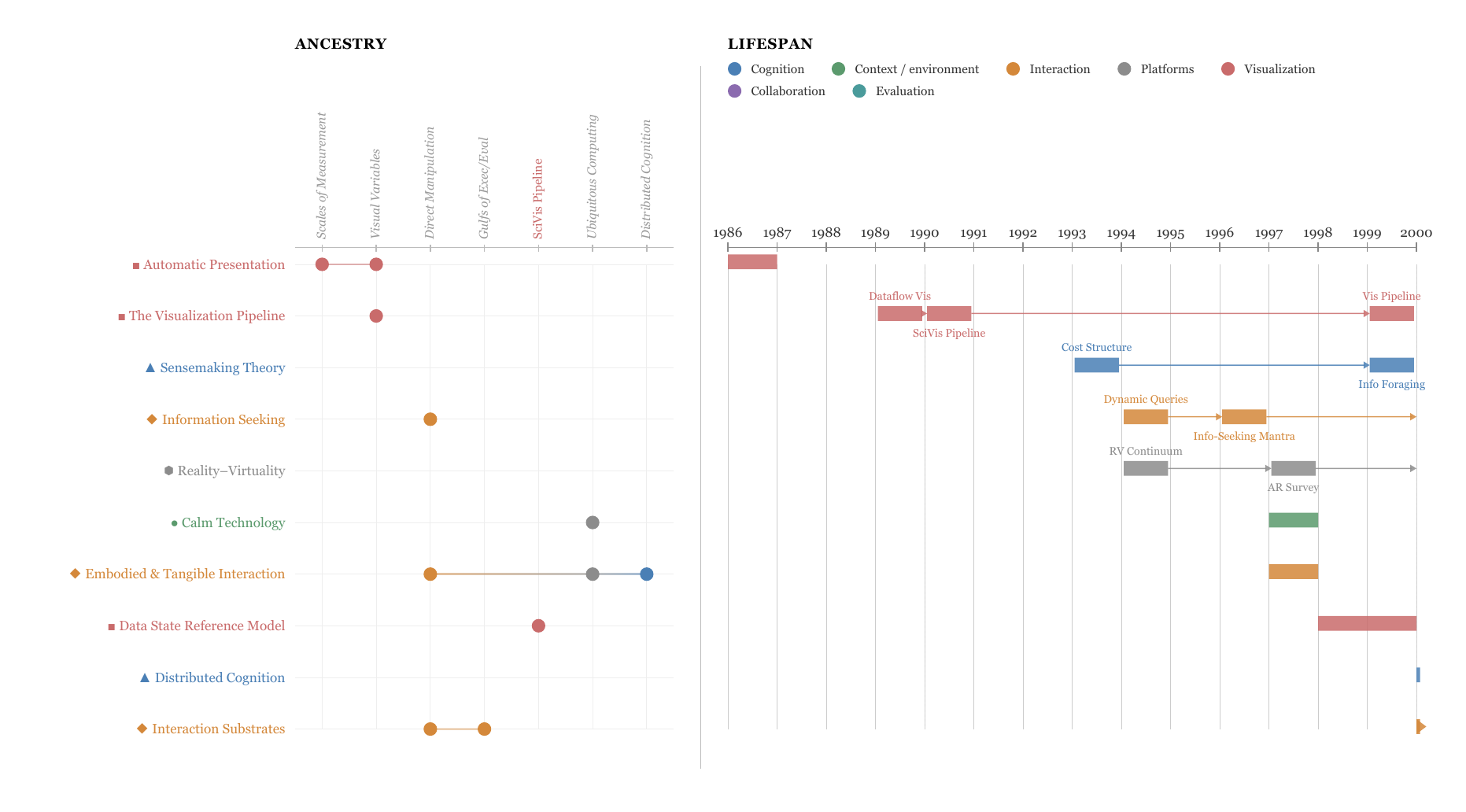}
    \caption{\textbf{Pre-2000 lineage.}
    Foundational work for ubiquitous analytics.
    Three families dominate the roots:
    a visualization line from Bertin's visual variables and Stevens' scales of measurement through the reference pipeline;
    an interaction line from Shneiderman's direct manipulation and Norman's gulfs;
    and a cognition line from Hutchins' distributed cognition.
    Sutherland's reality–virtuality lineage and Weiser's ubiquitous computing seed the platform vocabulary.
    }
    \label{fig:lineage-roots}
\end{figure*}

\section{Lineage and Research Themes}
\label{sec:themes}

Ubiquitous analytics~\cite{Elmqvist2013} was built on several pre-existing ideas: sensemaking, cognition, and data visualization itself, as well as more exotic ideas such as putting data in space, treating the body as an input device, and distributing thinking across artifacts.
Here, we trace this genealogy to reveal seven recurring themes that run the length of the field; we later use them to organize the rest of the paper.

\subsection{Roots: Before 2000}

The roots predate the field by decades, and they fall into three families that future work keeps drawing on (Figure~\ref{fig:lineage-roots}).

The \textbf{visualization} family starts with Bertin's retinal variables~\cite{Bertin1967} and Stevens' scales of measurement~\cite{Stevens1946}---a vocabulary for graphical marks and a typology of data---which together license Cleveland and McGill's empirical ranking of encodings~\cite{Cleveland1984} and Mackinlay's automatic presentation~\cite{Mackinlay1986}.
By the close of the century, this has yielded the reference pipeline~\cite{Card1999, DBLP:conf/infovis/Chi00}: data tables to visual structures to views.

The \textbf{interaction} family begins with Shneiderman's direct manipulation~\cite{DBLP:journals/computer/Shneiderman83} and Norman's gulfs of execution and evaluation~\cite{Norman1986}---continuous reversible action, and the two gaps that any interface has to bridge.
Dynamic queries and the information-seeking mantra~\cite{Shneiderman1994, Shneiderman1996} carry these into visualization, while Beaudouin-Lafon's instrumental interaction~\cite{Beaudouin-Lafon2000, Beaudouin-Lafon2004} reframes the goal from interfaces to interaction itself, a move whose descendants reach all the way to the substrates work of the 2020s~\cite{Mackay2025substrates}.

The \textbf{cognition} family is the one that matters most for our argument.
Hutchins' distributed cognition~\cite{Hutchins1995} holds that thinking extends beyond the skull to tools, artifacts, and other people; Kirsh's intelligent use of space and Scaife and Rogers' external cognition~\cite{DBLP:journals/ai/Kirsh95, Scaife1996} show how arranging the world offloads internal computation; and Hollan, Hutchins, and Kirsh~\cite{DBLP:journals/tochi/HollanHK00} install the whole edifice as a foundation for HCI.
Running alongside is the activity-theory strand---Vygotsky through Nardi to Kaptelinin~\cite{Vygotsky1978, Nardi1996, Kaptelinin2009}---and the sensemaking strand opened by Pirolli and Card's information foraging~\cite{Pirolli99}.

Two more roots sit apart but turn out to be foundational for our purposes.
Sutherland's ultimate display~\cite{sutherland1965ultimate}, by way of Milgram and Kishino's reality--virtuality continuum and Azuma's AR survey~\cite{Milgram1994, azuma97AR}, gives the field its \textbf{platform} vocabulary; Weiser's ubiquitous computing~\cite{Weiser1991} gives it its name and anytime/anywhere ambition.

\begin{figure*}[htb]
    \centering
    \includegraphics[width=\linewidth]{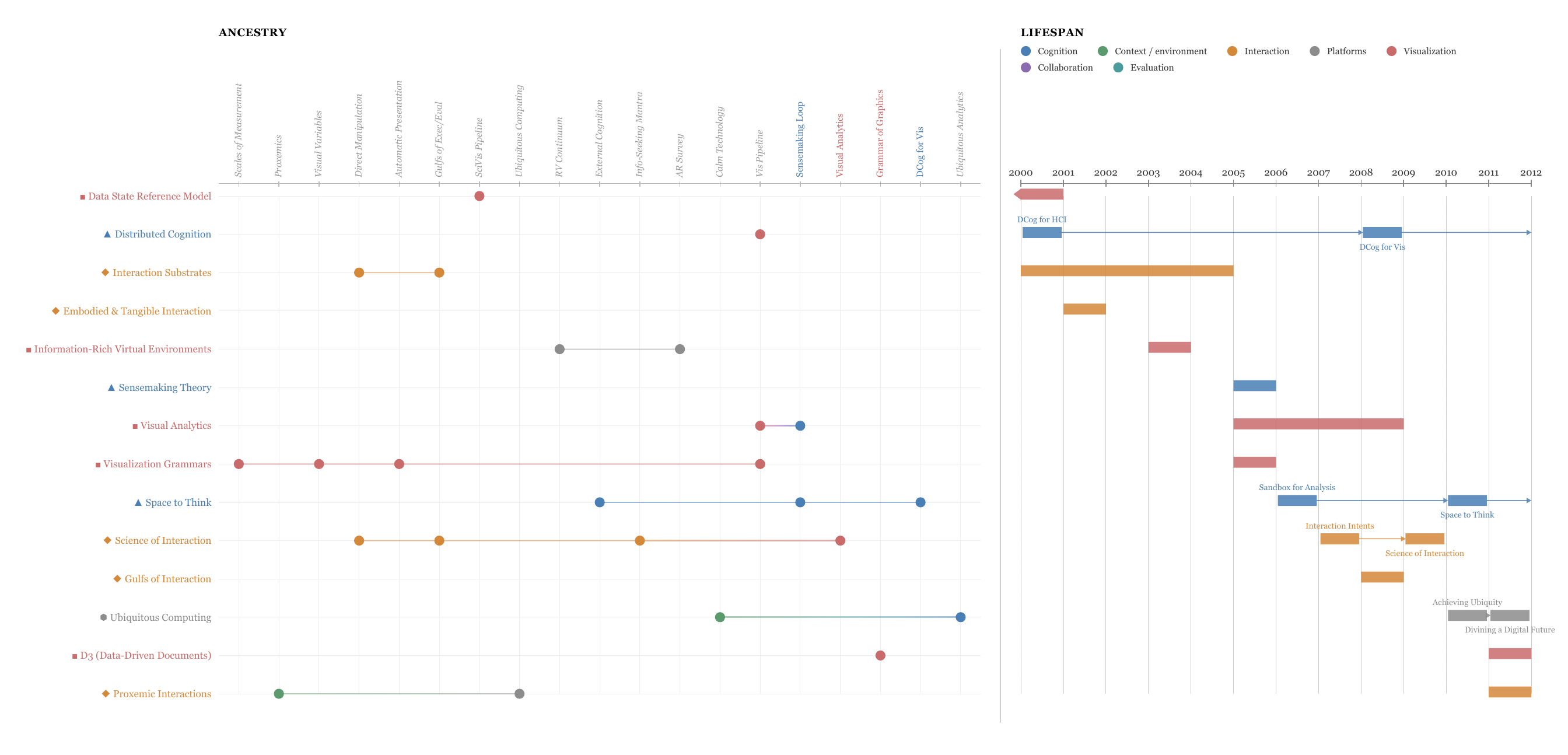}
    \caption{\textbf{2000--2012 lineage.}
    The second period, in which the visualization family consolidates and the cognition family turns toward analysis.
    Visual analytics arrives; visualization grammars take hold; interaction gains its own theory.
    The thread that matters most for the present agenda is the cognition spine running from distributed cognition for visualization through the Sandbox to Space to Think.
    }
    \label{fig:lineage-maturity}
\end{figure*}

\subsection{Visualization Matures: 2000--2012}

In this period the visualization family matures and the cognition family turns toward analysis (Figure~\ref{fig:lineage-maturity}).

Visual analytics arrives~\cite{Thomas2005, Keim2008}, fusing the reference pipeline with the sensemaking loop that Pirolli and Card had just formalized~\cite{Pirolli2005} on top of Russell et al.'s cost-structure~\cite{Russell1993}.
Interaction gets its own theory: Yi et al.'s intents and Pike et al.'s science of interaction~\cite{Yi2007, pike2009} argue that manipulating a view is not the same as conducting an analysis, and Lam identifies a third gulf---goal formation---beyond Norman's two~\cite{DBLP:journals/tvcg/Lam08}.
Grammars formalize the craft: Wilkinson's grammar of graphics~\cite{Wilkin2005GoG} through D3~\cite{Bostock2011} to Vega-Lite~\cite{Satyanarayan2017} yield a common declarative language.

A decisive event is Liu, Nersessian, and Stasko~\cite{Liu2008} applying distributed cognition to information visualization.
Their argument partly builds on Wright et al.'s Sandbox~\cite{Wright2006}, a freeform analytical workspace where evidence and hypotheses are arranged spatially, and from there into Andrews et al.'s Space to Think~\cite{Andrews2010} where analysts use large display space as external spatial memory for sensemaking.
That single line---external cognition~\cite{Scaife1996}, distributed cognition for vis, Sandbox, Space to Think---is the spine of everything that follows.
Meanwhile Bowman et al.'s information-rich virtual environments~\cite{DBLP:conf/vrst/BowmanNCPPY03} quietly seed the immersive turn, and proxemics enters HCI~\cite{Greenberg2011} on Hall's anthropological foundation~\cite{Hall1966}.

\begin{figure*}[htb]
    \centering
    \includegraphics[width=\linewidth]{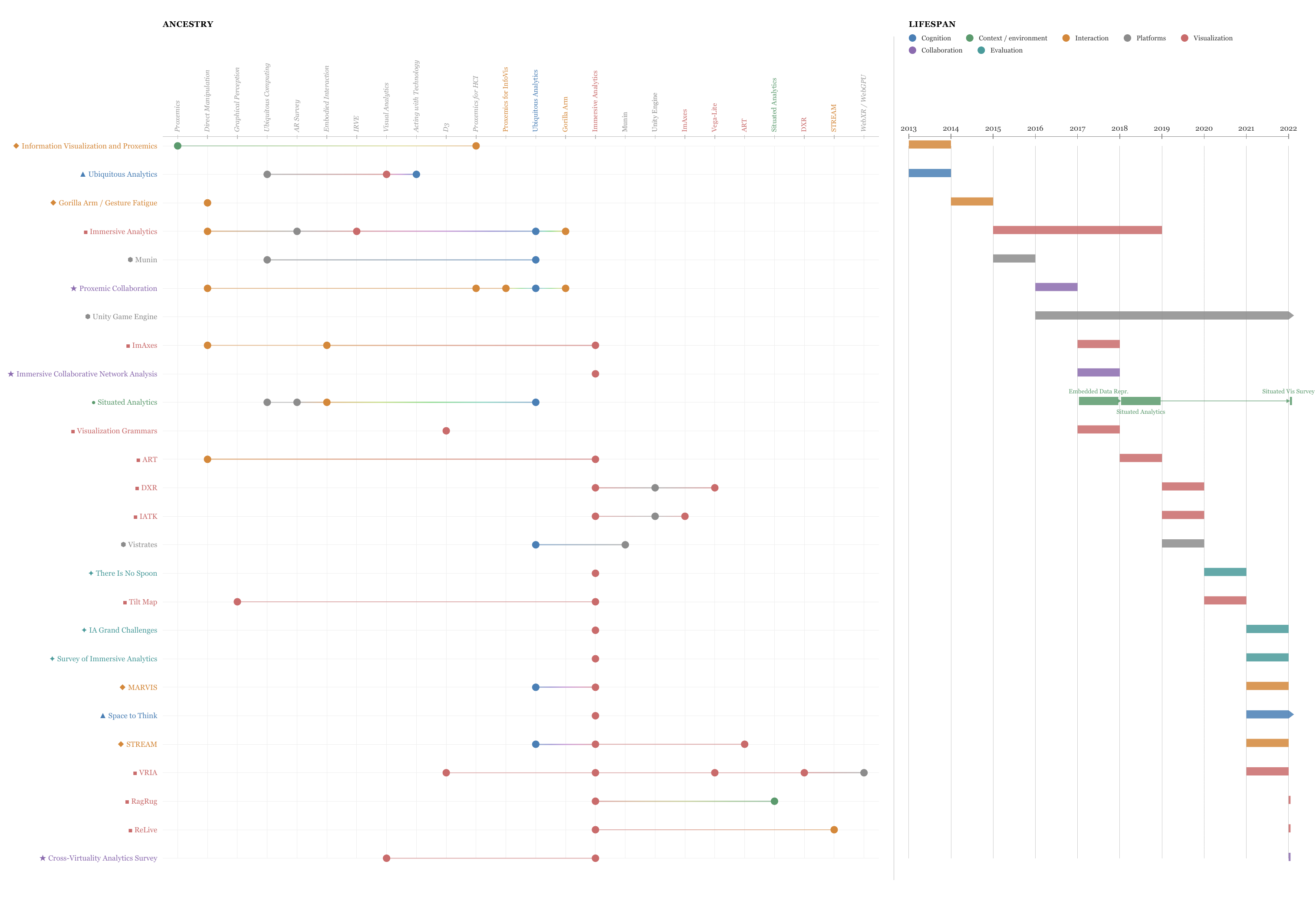}
    \caption{\textbf{2013--2022 lineage.}
    The decade ubiquitous analytics names itself and then splits into the subcommunities the field still works in.
    Ubiquitous analytics is coined; immersive analytics consolidates as a sibling; situated analytics splits off.
    The platform substrate matters here, with the engine-native toolkits built on Unity and VRIA opening the web-versus-engine split.
    The period closes with the first attempts to take stock, Ens et al.'s grand challenges among them.
    }
    \label{fig:lineage-emergence}
\end{figure*}

\subsection{Emergence: 2013--2022}

This is the decade the field names itself, and then fragments into the subcommunities within which still work (Figure~\ref{fig:lineage-emergence}).

Ubiquitous analytics is coined~\cite{Elmqvist2013}, drawing together ubicomp, visual analytics, and distributed cognition into a research program for analytics beyond the desktop.
Immersive analytics consolidates as a sibling~\cite{Chandler2015, Marriott2018}.
Situated analytics splits off along its own line, running from embedded data representations to situated analytics proper~\cite{willett17embedded, Thomas2018}.
Bressa et al.~\cite{Bressa2022} closes the line by disentangling what ``situated'' even means.
The platform substrate matters here: Unity becomes the default research engine, and the immersive toolkits---DXR, IATK---are built on it~\cite{sicat19dxr, cordeil19iatk}, while VRIA~\cite{butcher20vria} is proposed, based on open web technologies in the grand tradition of the visualization field.
Several systems are also introduced, including Munin~\cite{Badam2015}, Vistrates~\cite{Badam2019}, and Proxemic Lenses~\cite{Badam2016b}. 

The cognition theme extends cleanly: Lisle et al.'s Immersive Space to Think~\cite{DBLP:conf/vr/LisleDG0B21, DBLP:journals/tvcg/DavidsonLWBN23} carries Andrews et al.\ into VR and watches analysts shift from egocentric to linear layouts as schemas form.
And the period closes with the first attempts to take stock: the IA design-space reviews, the surveys, and Ens et al.'s grand challenges~\cite{Ens2021}.

\begin{figure*}[htb]
    \centering
    \includegraphics[width=\linewidth]{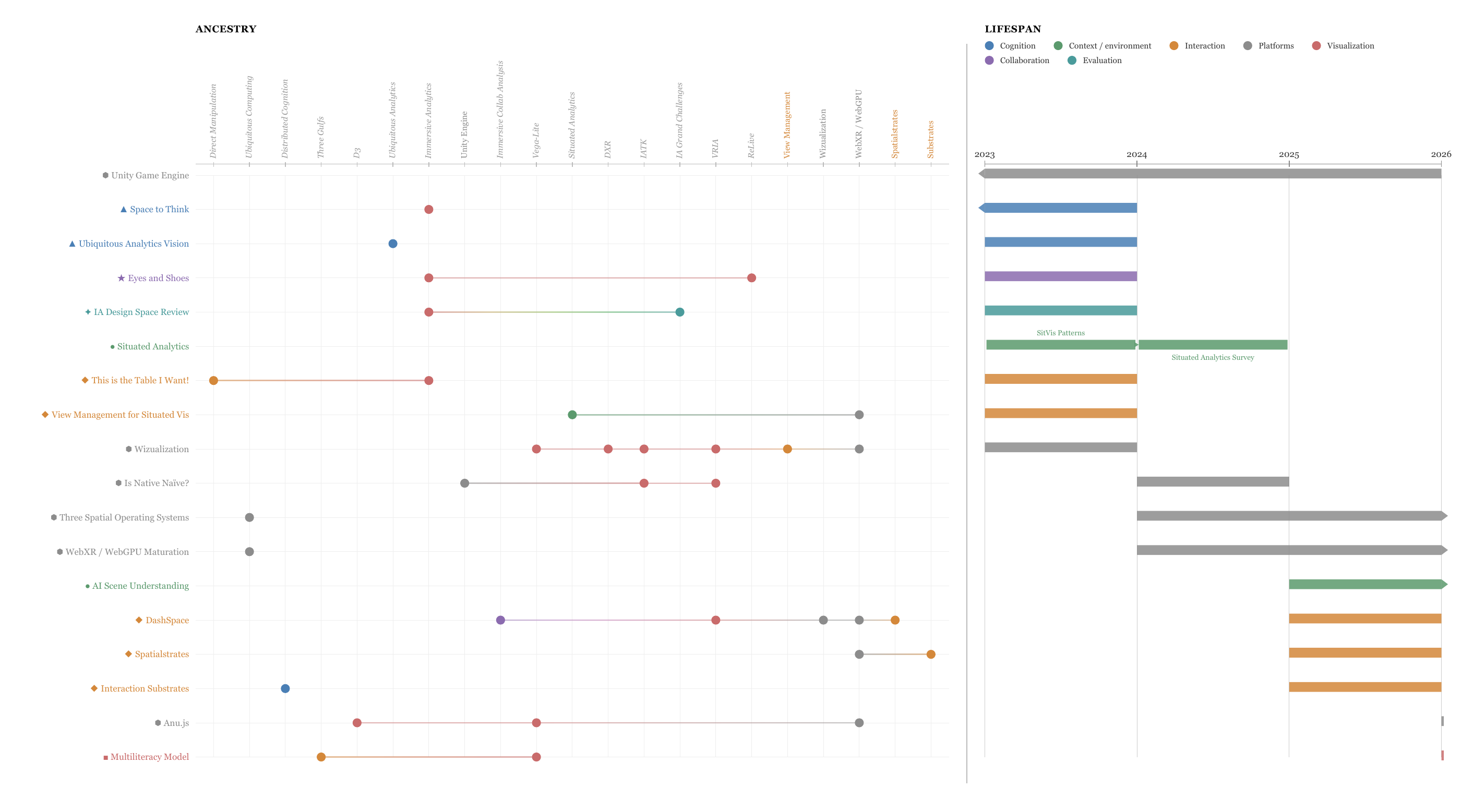}
    \caption{\textbf{2022-2026 lineage.}
    The most recent period where theory consolidates and the platform ground shifts.
    Data analytics as an anywhere and everywhere paradigm restates UA as a program; interaction substrates supply the term the framework runs on; DashSpace and Spatialstrates turn substrates into working platforms on open web standards.
    }
    \label{fig:lineage-theory}
\end{figure*}

\subsection{Toward Theory: 2022--2026}

The field has in recent years struggled towards consolidation (Figure~\ref{fig:lineage-theory}).
Two things happen at once: the theory consolidates, and platforms shift under everyone's feet.

On the theory side, ubiquitous analytics is revisited~\cite{Elmqvist2023}, and Mackay et al.'s interaction substrates~\cite{Mackay2025substrates}---structured media over which information is represented and propagated, descended from instrumental interaction and distributed cognition both---give us the term our framework runs on.
DashSpace and Spatialstrates~\cite{Borowski2025dashspace, Borowski2025spatialstrates} turn substrates into collaborative platforms on open web standards.

On the platform side, the change is structural rather than conceptual:
three spatial operating systems ship at once, AI scene understanding arrives as a platform service that makes the physical world queryable, and WebXR with WebGPU close the gap with proprietary engines~\cite{DBLP:journals/cga/ButcherBSMER24}.
Butcher et al.\ ask the question directly---is native naïve?---and the web-versus-engine line that opened with VRIA finally has evidence on both sides~\cite{DBLP:journals/cga/ButcherBSMER24}.
This is the convergence the introduction promised, drawn here as four bars running off the right edge of the timeline.

\subsection{Identifying Research Themes}
\label{sec:identification}

We coded each theme into one of seven clusters---cognition, visualization, interaction, platforms, context, collaboration, and evaluation---the categories the field already uses to organize itself.
Each theme gets the one cluster its main contribution belongs to.
Descent mostly stays within a cluster, so the columns read as vertical bands; the cross-cluster inheritances are the exceptions.
These are the themes the rest of the paper takes positions on, and we name them here in the order we argue them: from the analyst's mind outward to the infrastructure beneath.

\mypara{{\color{cog}\faBrain}~Cognition and the Analyst (Cog).}
How analysts think with the artifacts, devices, and spaces around them, modeled through post-cognitive frameworks.
The line runs from Hutchins' distributed cognition~\cite{Hutchins1995} through its application to HCI~\cite{DBLP:journals/tochi/HollanHK00} and to information visualization~\cite{Liu2008}, with the Space to Think studies~\cite{Andrews2010} as the empirical anchor.

\mypara{{\color{ctx}\faMapMarker}~With the World as My Context (Ctx).}
Analytics whose data is tied to physical places and objects, and whose environment the system can sense and interpret.
Embedded data representations~\cite{willett17embedded} and situated analytics~\cite{Thomas2018, Bressa2022} register data to the world; AI scene understanding makes that world legible to the system.

\mypara{{\color{int}\faHandPointer}~Interaction Beyond the Prototype (Int).}
The input side of analysis: how analysts act on data through gesture, gaze, voice, touch, and locomotion.
Direct manipulation~\cite{DBLP:journals/computer/Shneiderman83} and Norman's gulfs~\cite{Norman1986} sit at the root, with modality-specific work on proxemics, gesture fatigue~\cite{HincapieRamos2014}, and hybrid surfaces layered on top.

\mypara{{\color{plat}\faLayerGroup}~Platforms and the Open Standards Race (Platform).}
The infrastructure analytics runs on: engines, toolkits, devices, and the web stack beneath them.
Toolkits split into engine-native (Unity, DXR~\cite{sicat19dxr}, IATK~\cite{cordeil19iatk}) and web-native (VRIA~\cite{butcher20vria}, Anu.js~\cite{Saffo_Anujs}, Spatialstrates~\cite{Borowski2025spatialstrates}) lines, against a backdrop of three spatial operating systems and maturing standards such as WebXR and WebGPU.

\mypara{{\color{vis}\faCube}~Visualization in Space (Vis).}
The design of visual representations for three-dimensional and immersive media.
Information-rich virtual environments~\cite{DBLP:conf/vrst/BowmanNCPPY03}, embodied axes~\cite{cordeil17imaxes}, the immersive toolkits, and the immersive visualization grammars make up the line.

\mypara{{\color{collab}\faUsers}~Anytime Anywhere Together (Collab).}
Analysis as collaborative work across people, devices, and degrees of co-location.
Immersive collaborative network analysis~\cite{cordeil17immersivecollab}, proxemic lenses~\cite{Badam2016b}, and the cross-virtuality and cross-reality frameworks~\cite{DBLP:journals/cgf/FrohlerAPFSRTHB22, DBLP:conf/chi/SaffoBDE23} populate it.

\mypara{{\color{eval}\faFlask}~Evaluation in the Wild (Eval).}
How the field measures whether any of this works.
The line is short---a few controlled studies~\cite{batch20econimmersive}, the immersive analytics surveys, and Ens et al.'s grand challenges~\cite{Ens2021}---because evaluation has mostly stayed in the laboratory.
 
\section{Future Challenges}
\label{sec:challenges}

Table~\ref{tab:challenge-matrix} gives an overview of the research challenges resulting from pairwise convergence described in Section~\ref{sec:method}.
In the following subsections, we discuss each theme in turn to present the research challenge in more detail, including a potential deliverable arising from each.

\begin{table*}[htbp]
    \centering
    \caption{\textbf{Research challenge matrix.}
      Rows are the \emph{parent} theme; columns are the \emph{partner} theme that inflects the challenge.
      Each cell names the most fundamental open question at that intersection.
      Order matters, so the matrix is not symmetric.
      }
    \label{tab:challenge-matrix}
    \small
    \SetTblrInner{rowsep=3pt,colsep=4pt}
    \begin{tblr}{
      colspec = {l|c|c|c|c|c|c|c},
      row{1} = {font=\bfseries\small, bg=gray!10, halign=c},
      column{1} = {font=\bfseries\small, bg=gray!10, halign=l},
      cell{1}{1} = {bg=gray!20},
      hline{1,9} = {0.8pt, solid},        
      hline{2} = {0.8pt, solid},           
      vline{1,9} = {0.8pt, solid},         
      vline{2} = {0.8pt, solid},           
      hline{3,4,5,6,7,8} = {0.3pt, solid}, 
      vline{3,4,5,6,7,8} = {0.3pt, solid}, 
      cells = {halign=c, valign=m, font=\scriptsize},
    }
      {\small\textbf{Parent$\,\downarrow$}}
      & {\color{cog}\faBrain}~Cog
      & {\color{ctx}\faMapMarker*}~Ctx
      & {\color{int}\faHandPointer}~Int
      & {\color{plat}\faLayerGroup}~Plat
      & {\color{vis}\faCube}~Vis
      & {\color{collab}\faUsers}~Collab
      & {\color{eval}\faFlask}~Eval \\

      {\color{cog}\faBrain}~\textbf{Cog}
      & {---}
      & {Space, place,\\and memory}
      & {Affordance and\\mental models}
      & {Cognitive\\guarantees}
      & {Schema\\from space}
      & {Shared\\schemas}
      & {Measuring cognition\\in the field} \\

      {\color{ctx}\faMapMarker*}~\textbf{Ctx}
      & {Environmental\\cognitive load}
      & {---}
      & {Physical-virtual\\interplay}
      & {World\\anchoring}
      & {Embedded data\\representation}
      & {Context sharing\\and adaptation}
      & {In situ\\evaluation} \\

      {\color{int}\faHandPointer}~\textbf{Int}
      & {Channel cognitive\\properties}
      & {Spatial\\manipulation}
      & {---}
      & {Cross-device\\channel translation}
      & {Spatial vis\\interaction}
      & {Team-first\\interaction}
      & {Channel ranking\\and literacy} \\

      {\color{plat}\faLayerGroup}~\textbf{Plat}
      & {Platform capability\\envelope}
      & {Environmental\\knowledge}
      & {Cross-reality and\\cross-device continuity}
      & {---}
      & {Visual fidelity\\and scale}
      & {Collaboration \&\\shared state}
      & {Integrated eval\\infrastructure} \\

      {\color{vis}\faCube}~\textbf{Vis}
      & {Spatial encoding\\perception}
      & {Embedding and\\its limits}
      & {Spatial interaction\\idioms}
      & {Spatial vis\\grammars}
      & {---}
      & {Specialized vs.\\shared views}
      & {Graphical perception\\in the wild} \\

      {\color{collab}\faUsers}~\textbf{Collab}
      & {Leveraging\\multiple minds}
      & {Proxemics in\\spatial analytics}
      & {Asymmetric device\\collaboration}
      & {Collaboration \\across platforms}
      & {View negotiation\\and access}
      & {---}
      & {Beyond \\ collaborative eval} \\

      {\color{eval}\faFlask}~\textbf{Eval}
      & {Longitudinal\\cognitive capture}
      & {Measuring real-world\\cognitive load}
      & {Sensor fusion for\\interaction logging}
      & {Spatial evaluation\\platforms}
      & {Visualizing\\evaluation data}
      & {Collaborative\\evaluation practice}
      & {---} \\

    \end{tblr}
\end{table*}

\subsection[Cognition and the Analyst]{{\color{cog}\faBrain}~Cognition and the Analyst}
\label{sec:ch-cog}

If sensemaking is distributed across the analyst and their surroundings, then the analyst's mind is the right place to start.
The six challenges below take cognition as the parent and ask what each of the other themes does to it: what the environment costs, what input demands, what a platform must guarantee, what space does to a schema, how cognition spreads across people, and how any of it can be measured once the lab door is open.

\mypara{{\color{ctx}\faMapMarker*}~Space, place, and memory.}
How do analysts use spatial memory, and the distinction between geometric space and meaningful place, to organize and recover analytical work?
\textit{Deliverable:} empirical evidence on whether spatial memory for analytical content, not just for navigation, supports schema recovery across sessions and environments.

\mypara{{\color{int}\faHandPointer}~Affordance and mental models.}
How do analysts form mental models of spatial input affordances, and what are the cognitive costs of different input channels for analytical tasks?
\textit{Deliverable:} a cognitive characterization of spatial input channels, mapping each to the mental model an analyst forms about what is interactable and how.

\mypara{{\color{plat}\faLayerGroup}~Cognitive guarantees.}
What higher-level cognitive support, such as persistence, undo, state recovery, and analytical history, must a platform provide for distributed analytical reasoning?
\textit{Deliverable:} a minimum specification of platform-level cognitive guarantees that analytical applications can rely on across devices.

\mypara{{\color{vis}\faCube}~Schema from space.}
How do spatial and embedded visual encodings affect schema construction differently from screen-based ones?
\textit{Deliverable:} comparative studies linking encoding type (2D screen, spatial floating, world-embedded) to schema complexity and analytical depth.

\mypara{{\color{collab}\faUsers}~Shared schemas.}
How is analytical cognition distributed across multiple analysts working on shared substrates?
\textit{Deliverable:} field observations of distributed analytical cognition that document how schemas propagate, merge, or conflict across collaborating analysts.

\mypara{{\color{eval}\faFlask}~Measuring cognition in the field.}
How do we measure distributed analytical cognition in the field rather than in the lab?
\textit{Deliverable:} validated observational methods and metrics for capturing schema construction, substrate switching, and cognitive distribution in deployed systems.

\subsection[With the World as My Context]{{\color{ctx}\faMapMarker*}~With the World as My Context}
\label{sec:ch-ctx}

The environment is no longer scenery behind the screen; once AI can read a room, the room is part of the computation.
Taking context as the parent, we here explore what the world does to the other themes.

\mypara{{\color{cog}\faBrain}~Environmental cognitive load.}
How does the physical environment constrain, distract from, or impose on the analyst's cognitive work?
\textit{Deliverable:} models of environmental cognitive load that predict when real-world settings help versus hinder analytical reasoning.

\mypara{{\color{int}\faHandPointer}~Physical-virtual interplay.}
How does the interplay between physical and virtual shape what interactions are possible during situated analysis?
\textit{Deliverable:} a characterization of how environmental properties (surfaces, objects, lighting, noise) constrain and afford interaction techniques for analytical tasks.

\mypara{{\color{plat}\faLayerGroup}~World anchoring.}
How do platforms anchor analytical artifacts to the physical world and hold that anchoring over time?
\textit{Deliverable:} design patterns for spatial registration and persistence of analytical content across sessions, including requirements for reference spaces and relocalization.

\mypara{{\color{vis}\faCube}~Embedded data representation.}
How should data sensed from the physical world be visualized back into it, and what are the design principles for that round-trip?
\textit{Deliverable:} design guidelines for embedded data representations that account for the full cycle from physical sensing to situated visualization.

\mypara{{\color{collab}\faUsers}~Context sharing and adaptation.}
How do collaborators in different physical environments make their respective contexts legible to each other?
\textit{Deliverable:} techniques for adapting one collaborator's physical context to another, enabling cross-context analytical collaboration.

\mypara{{\color{eval}\faFlask}~In situ evaluation.}
How do we design valid evaluations when a system's analytical value depends on the specific physical environment it inhabits?
\textit{Deliverable:} evaluation methodologies for situated systems that account for site-dependence without requiring laboratory reproduction of the deployment environment.

\subsection[Interaction Beyond the Prototype]{{\color{int}\faHandPointer}~Interaction Beyond the Prototype}
\label{sec:ch-int}

Interaction techniques that work in a demo often fail the moment a hand shakes, a room is loud, or the device in play is not the one the technique was designed for.
Starting with interaction, the six challenges ask what each theme demands of input once it has to survive contact with the world. 

\mypara{{\color{cog}\faBrain}~Channel cognitive properties.}
What role does physical interaction play as knowledge construction, and how do spatial input channels differ in their cognitive properties?
\textit{Deliverable:} an experimentally grounded characterization of spatial input channels by their cognitive properties, beyond throughput and error rate, including their role in analytical reasoning.

\mypara{{\color{ctx}\faMapMarker*}~Spatial manipulation.}
How do analysts directly manipulate data that is situated in, or registered to, the physical world?
\textit{Deliverable:} interaction techniques for spatial manipulation of world-registered analytical content, with empirical evidence of their effectiveness in situ.

\mypara{{\color{plat}\faLayerGroup}~Cross-device channel translation.}
How do interaction vocabularies translate between devices and platforms without losing analytical meaning?
\textit{Deliverable:} translation mappings for interaction techniques across device classes (phone, tablet, headset, wall display), identifying where meaning is preserved and where it breaks.

\mypara{{\color{vis}\faCube}~Spatial vis interaction.}
What interaction techniques support creating, manipulating, and annotating spatial and situated visualizations?
\textit{Deliverable:} a catalog of spatial visualization interaction techniques with empirical evidence for their analytical utility.

\mypara{{\color{collab}\faUsers}~Team-first interaction.}
How do we design interaction techniques that prioritize the team's collective analytical process over individual workflows; e.g., team-first~\cite{Badam2017c}?
\textit{Deliverable:} interaction designs and evaluations where team-level analytical coordination is the primary design objective.

\mypara{{\color{eval}\faFlask}~Channel ranking and literacy.}
Can we rank spatial input channels for expressiveness and effectiveness, and how do we measure analysts' interactive literacy for spatial media?
\textit{Deliverable:} a Mackinlay-style ranking~\cite{Mackinlay1986} of spatial input channels for analytical tasks, with validated instruments for measuring interactive visualization literacy~\cite{DBLP:conf/chi/LeonBVE26} in space.

\subsection[Platforms and the Open Standards Race]{{\color{plat}\faLayerGroup}~Platforms and the Open Standards Race}
\label{sec:ch-plat}

Three spatial operating systems are now shipping, and the decisions their vendors make this decade will harden into the platform we are stuck with.
Extracting challenges based on platforms asks what the infrastructure owes the layers above it: a cognitive floor, environmental knowledge, continuity across reality levels, rendering headroom, collaboration primitives, and instrumentation.

\mypara{{\color{cog}\faBrain}~Platform capability envelope.}
How do platform capabilities such as field of view, refresh rate, and tracking accuracy shape what analytical reasoning is possible?
\textit{Deliverable:} empirical mappings from platform capability parameters to analytical task performance, identifying thresholds below which specific reasoning tasks become impractical.

\mypara{{\color{ctx}\faMapMarker*}~Environmental knowledge.}
How should platforms expose environmental data, such as scene graphs, object recognition, and spatial persistence, to analytical applications?
\textit{Deliverable:} API design patterns and reference implementations for exposing environmental knowledge to analytical applications across spatial operating systems.

\mypara{{\color{int}\faHandPointer}~Cross-reality and cross-device continuity.}
How do analytical workflows span devices, form factors, and reality levels without breaking the analyst's task flow?
\textit{Deliverable:} cross-reality architectural patterns and reference implementations that support continuous analytical workflows across heterogeneous devices.

\mypara{{\color{vis}\faCube}~Visual fidelity and scale.}
What display constraints (resolution, refresh rate, ambient light, dataset scalability) do platforms impose on visualization, and where are the bottlenecks?
\textit{Deliverable:} systematic benchmarks of visualization performance across current spatial platforms, identifying where open web standards~\cite{DBLP:journals/cga/ButcherBSMER24} (WebXR, WebGPU) fall short and where they suffice.

\mypara{{\color{collab}\faUsers}~Collaboration beyond shared state.}
How do platforms support role-differentiated, device-asymmetric collaboration beyond simple state synchronization?
\textit{Deliverable:} platform mechanisms for collaborative sessions that support differentiated roles, views, and capabilities across heterogeneous devices.

\mypara{{\color{eval}\faFlask}~Integrated evaluation infrastructure.}
How do we embed evaluation instrumentation and crowdsourced study protocols into spatial platforms?
\textit{Deliverable:} a spatial extension of evaluation frameworks such as ReVISit~\cite{DBLP:journals/tvcg/CutlerWSDBNHMHL26} that supports study design, execution, and crowdsourced data collection for ubiquitous analytical systems.

\subsection[Visualization in Space]{{\color{vis}\faCube}~Visualization in Space}
\label{sec:ch-vis}

Spatial visualization design is still in its infancy; we have decades of theory for marks on a 2D plane and almost none for marks suspended in a room or pinned to an object.
Here we explore what each theme demands of the encoding itself: perception, the limits of embedding, interaction idioms, a grammar, the private--public boundary, and perception measured in the wild.

\mypara{{\color{cog}\faBrain}~Spatial encoding perception.}
How do spatial and volumetric visual encodings compare to established 2D conventions for perceptual accuracy and task completion time?
\textit{Deliverable:} a body of perceptual studies extending the Cleveland-McGill tradition~\cite{Cleveland1984} to spatial and volumetric encodings under both laboratory and real-world conditions.

\mypara{{\color{ctx}\faMapMarker*}~Embedding and its limits.}
When should data be embedded in its physical referent, and when does embedding hinder rather than help understanding?
\textit{Deliverable:} empirical guidelines specifying the conditions (data type, task, environment) under which embedded visualization~\cite{willett17embedded} outperforms detached alternatives, and when it does not.

\mypara{{\color{int}\faHandPointer}~Spatial interaction idioms.}
What are the spatial equivalents of the standard visualization interaction idioms: pan, zoom, filter, details-on-demand?
\textit{Deliverable:} a catalog of spatial interaction idioms for visualization, empirically validated for analytical utility in spatial computing environments.

\mypara{{\color{plat}\faLayerGroup}~Spatial visualization grammars.}
What would a declarative visualization grammar for spatial computing look like: the spatial Vega-Lite?
\textit{Deliverable:} a declarative grammar and authoring environment for  spatial visualizations across platforms, with abstractions suitable for domain experts.

\mypara{{\color{collab}\faUsers}~Specialized vs.\ shared views.}
How should visual representations differentiate an analyst's specialized private view from the shared collaborative one?
\textit{Deliverable:} design patterns for managing the private--public visualization boundary in collaborative spatial analytical environments.

\mypara{{\color{eval}\faFlask}~Graphical perception in the wild.}
How do we measure graphical perception for spatial encodings under real-world conditions rather than controlled laboratory settings?
\textit{Deliverable:} perceptual benchmarks and experimental protocols for evaluating spatial visual encodings in situ, accounting for environmental variability.

\subsection[Anytime Anywhere Together]{{\color{collab}\faUsers}~Anytime Anywhere Together}
\label{sec:ch-collab}

Analytical work is social~\cite{Heer2007}, and in ubiquitous analytics settings, collaborators rarely have matching hardware: one wears a headset, another holds a phone, a third is at a wall display across the building.
Eliciting challenges from the viewpoint of collaboration involves studying how each theme bears on working together under asymmetry: multiple minds, proxemics, asymmetric input, cross-platform mechanics, view negotiation, and evaluating joint outcomes.

\mypara{{\color{cog}\faBrain}~Leveraging multiple minds.}
How do we best leverage multiple analysts as a distributed cognitive system for sensemaking?
\textit{Deliverable:} DCog-grounded design frameworks for collaborative systems that distribute cognitive work well across people.

\mypara{{\color{ctx}\faMapMarker*}~Proxemics in spatial analytics.}
How do proxemics, physical or avatar-mediated, function as implicit coordination channels in spatial analytical environments?
\textit{Deliverable:} empirical evidence on how spatial relationships between analysts (physical or virtual) affect coordination, awareness, and analytical outcomes.

\mypara{{\color{int}\faHandPointer}~Asymmetric device collaboration.}
How do asymmetric input capabilities across collaborators affect collaborative analytical workflows?
\textit{Deliverable:} empirical studies and design guidelines for collaborative tasks where participants use devices with fundamentally different input capabilities.

\mypara{{\color{plat}\faLayerGroup}~Collaboration across platforms.}
How do we enable collaboration when participants have different platforms, devices, and capability?
\textit{Deliverable:} collaboration architectures and protocols that work across heterogeneous platforms without requiring a shared hardware or software stack.

\mypara{{\color{vis}\faCube}~View negotiation and access.}
How do collaborators negotiate what data is visible, to whom, and at what level of detail in a shared environment?
\textit{Deliverable:} mechanisms for dynamic view negotiation and access control in collaborative spatial analytical environments.

\mypara{{\color{eval}\faFlask}~Beyond collaborative evaluation.}
How do we step past current collaborative-visualization evaluation methods to capture the quality of a team's analytical outcome?
\textit{Deliverable:} evaluation methods and metrics that assess collaborative products (shared schemas, joint decisions, integrated reports) rather than individual task performance alone.

\subsection[Evaluation in the Wild]{{\color{eval}\faFlask}~Evaluation in the Wild}
\label{sec:ch-eval}

Evaluation has to leave the laboratory, because a ubiquitous analytics system that only works under controlled conditions has not been tested where it lives.
Here we study how themes reshape methods: longitudinal cognition capture, real-world load, sensor fusion, study platforms, visualizing study data itself, and evaluation as shared practice.

\mypara{{\color{cog}\faBrain}~Longitudinal cognitive capture.}
How do how analysts' cognitive strategies evolve as they adopt spatial tools?
\textit{Deliverable:} longitudinal study protocols and validated instruments for tracking cognitive adaptation to environments.

\mypara{{\color{ctx}\faMapMarker*}~Measuring real-world cognitive load.}
How do we measure the cognitive load that real-world environments impose on analytical tasks in specific settings?
\textit{Deliverable:} context-sensitive cognitive load measurement techniques validated for situated analytical work in real environments.

\mypara{{\color{int}\faHandPointer}~Sensor fusion for interaction logging.}
How do we fuse data from gaze, gesture, movement, and physiological sensors to capture spatial analytical interaction?
\textit{Deliverable:} standardized multi-sensor logging frameworks and analysis pipelines for spatial analytical interaction data.

\mypara{{\color{plat}\faLayerGroup}~Spatial evaluation platforms}
How do we extend evaluation frameworks like ReVISit to support study design, execution, and crowdsourced data collection for spatial analytical systems?
\textit{Deliverable:} a spatial computing extension of ReVISit~\cite{DBLP:journals/tvcg/CutlerWSDBNHMHL26} (or equivalent) for designing, deploying, and analyzing user studies of ubiquitous analytical systems.

\mypara{{\color{vis}\faCube}~Visualizing evaluation data}
How do we visualize the results of spatial analytical studies, extending approaches like ReLive to make the study data itself analytically tractable?
\textit{Deliverable:} visualization tools purpose-built for exploring and communicating the results of spatial analytics user studies (such as ReLive~\cite{DBLP:conf/chi/HubenschmidWFBZ22}).

\mypara{{\color{collab}\faUsers}~Collaborative evaluation practice.}
How do we enable collaboration in evaluation itself: designing, running, and analyzing experiments as a shared research practice?
\textit{Deliverable:} collaborative evaluation platforms and workflows that support multi-site, multi-researcher experimental practice for ubiquitous analytics.

\section{Discussion}
\label{sec:discussion}

We present 42 research challenges for ubiquitous analytics as the outcome of a deliberate approach where seven research themes from the literature are studied pairwise to understand potential future research topics.
In this section, we step back and ask what this method captures, what it misses, and what cross-cutting threads remain unexplored.

\subsection{Convergence as Method}

The generative principle behind the challenge matrix is simple: pair every theme with every other, let the first theme dominate, and ask what the second changes about it.
However, convergence privileges boundary problems: the questions that live between two themes.
It is less suited to surfacing deep intra-theme problems.
We do not claim our review is exhaustive, but that it is a lens, and like any lens it sharpens some things at the expense of others.

The alternative would have been to follow each theme individually and generate challenges from within.
That is what most research agendas do, including Ens et al.~\cite{Ens2021}.
The intra-theme approach produces challenges that are easier to scope as individual projects; the convergence approach produces challenges that are harder to scope but more likely to move the field, precisely because they refuse to stay inside one silo.
We chose convergence because the moment demands it: the forces now acting on ubiquitous analytics---spatial computing, AI, open standards---are interesting individually and transformative in combination.

\subsection{Relationship to Prior Challenges}

Ens et al.'s 17 challenges, organized under four themes (spatially situated visualization, interaction, collaborative analytics, and scenarios and evaluation), were produced by 24 experts through workshop consensus.
Our 42 challenges, organized under seven themes, are produced through a position-driven analysis in 2026.
There are several overlaps and differences between our works.

\mypara{What we have in common.}
Many of Ens et al.'s challenges reappear in our matrix, reframed as convergence questions.
Their call for understanding spatial perception of data maps onto our Vis~$\times$~Cog (spatial encoding perception); their challenge on multi-user collaboration maps onto our Collab row; their concern with evaluation methodology maps onto our Eval column.
Where we differ is in how they are framed: as boundary questions between two themes rather than open problems within one.

\mypara{What they have that we do not.}
Ens et al.\ give more attention to the phenomenology of immersion itself: presence, embodiment, and the felt experience of ``stepping through the glass.''
Our agenda, scoped to ubiquitous rather than immersive analytics alone, does not foreground immersion as a category.
A phone-based analyst has no sense of presence in the VR sense; the cognitive and interaction questions still apply.
This is a deliberate tradeoff: we gain scope and lose depth on the specifically immersive experience.

\mypara{What we have that they do not.}
Three themes in our matrix have no counterpart in Ens et al.: \textit{Platforms and the Open Standards Race}, \textit{With the World as My Context} (AI-mediated environmental understanding), and the explicit treatment of \textit{Cognition} as a theme grounded in distributed cognition rather than a background assumption.
The platform theme did not exist in 2020 because there was effectively one spatial OS; there are now three.
The context theme did not exist because AI scene understanding was not a platform service; it is now.
And the cognitive theme was implicit in Ens et al.\ but never stated as a position with an evidence gap.
These three themes alone generate 18 of our 42 challenges---nearly half the matrix---none of which could have appeared in the 2020 agenda.

\mypara{What the timing changes.}
The six years between the two agendas are not incremental.
Ens et al.\ wrote when spatial computing was a research-lab concern with one commercially viable headset (HoloLens 2) and no ecosystem competition.
We write when three operating systems ship, when slim-form-factor glasses sell in consumer electronics stores, and when generative AI has rewritten the assumptions about what a platform can sense, infer, and generate.
Several of their challenges---toolkit development, for instance---have been partially addressed by the ecosystem that has since emerged.
Others, like evaluation methodology, remain as open as ever.

\subsection{AI as a Cross-Cutting Force}

AI appears in 14 of the 42 challenge cells, more than any other single thread.
It is not a theme of its own in our matrix because its reach extends across multiple themes.

\mypara{Cognition.}
Agentic AI changes the distribution of cognition.
The analyst's sensemaking loop, traditionally distributed across human memory, external artifacts, and collaborators, now includes a non-human autonomous agent that can read the data, see the room, and take actions on spatial substrates.
The DCog framing handles this naturally---the agent is another node in the representational network---but the empirical questions are open: does the analyst's sense of agency change?
Does trust calibration follow the same dynamics as human-human collaboration, or something new?
When the AI constructs a schema on a substrate the analyst did not request, what happens to the analysis?

\mypara{Context.}
This is where AI is most visible today.
Platform-level scene understanding (Android XR's Gemini integration, visionOS's spatial awareness) turns the physical environment into a queryable substrate.
The research question is how to connect environmental knowledge to analytical reasoning, including overlaying data on surfaces as well as using what the AI knows about the room to filter, prioritize, and contextualize the analysis.

\mypara{Interaction.}
AI enables automatic channel adaptation: sensing environmental constraints (noise, light, social context) and shifting interaction channels without explicit user command.
The open question is reliability.
Graceful degradation in a controlled demo is not the same as graceful degradation on a factory floor.

\mypara{Platforms.}
Vibe coding---AI-generated spatial applications from natural language prompts---is structurally better positioned on web standards (WebXR, WebGPU) than on proprietary engines: a web application is a self-contained text artifact that an LLM can generate, run, and repair in one loop, whereas a game engine project entangles code with serialized scenes, assets, and editor state that resist LLM reasoning~\cite{Du2026}, and generation quality tracks the prevalence of a language in training corpora~\cite{DBLP:journals/tse/CassanoGNNPPYZAFGGJ23}, where web technologies dominate.
A vibe-coded analytical dashboard may look right and compute wrong; the field needs validity checks that fit into AI-assisted development workflows.

\mypara{Visualization.}
AI-assisted authoring could close the gap between the domain expert who has the data and the visualization designer who has the craft.
The authoring gap is wider for spatial visualization than it ever was for 2D; AI generation could narrow it.
But generated visualizations need to obey the same perceptual and cognitive principles as hand-crafted ones---expressiveness, effectiveness, and the design constraints a dedicated  grammar~\cite{DBLP:journals/tvcg/BatchBRE24} would encode.

\mypara{Collaboration.}
AI can facilitate mediation between collaborators on asymmetric devices: translating representations across platforms, summarizing analytical state for a newcomer, resolving conflicts between competing schemas.
And while AI can never be thought of as a collaborative partner on par with human actors~\cite{Shneiderman2022}, LLM-powered autonomous agents will likely have significant impact in future human-AI collaboration~\cite{DBLP:conf/chi/AmershiWVFNCSIB19}.

\mypara{Evaluation.}
AI generates data (interaction logs, gaze traces, physiological signals) at a volume that existing analysis methods cannot absorb.
It also introduces a new evaluation target: the AI agent itself.
How do we evaluate whether an AI cognitive partner actually improves analytical outcomes, rather than merely increasing the analyst's confidence?
This is a calibration problem that evaluation methodology has not yet addressed for spatial analytics.

\section{Conclusion}
\label{sec:conclusion}

In this paper, we have discussed the convergence of spatial computing and what it will mean for future research in ubiquitous analytics.
Using a careful review of the literature, we have presented seven positions on where ubiquitous analytics has been and where it is going.
By studying the intersection of these positions, we have further derived 42 research challenges from their pairwise convergence.
The positions are falsifiable, the challenges are scoped to be tractable as individual research projects, and the matrix that organizes them is a tool for seeing which problems the field is working on and which it is ignoring.
We look forward to seeing how the field will respond.

\section*{Acknowledgments}

This work was partially supported by Villum Investigator grant VL-54492 by Villum Fonden.
Any opinions, findings, and conclusions expressed in this material are those of the authors and do 
not necessarily reflect the views of the funding agency.

\section*{Disclosure of Generative AI Content}

The visualization software used to represent historical contributions (e.g., the lineage plots in Section~\ref{sec:themes}) were built with the help of Anthropic's Claude Opus 4.6 and 4.7.
Generative AI was furthermore used for editing and grammar enhancement of the article according to IEEE's \href{https://journals.ieeeauthorcenter.ieee.org/become-an-ieee-journal-author/publishing-ethics/guidelines-and-policies/submission-and-peer-review-policies/#ai-generated-content}{Guidelines for Artificial Intelligence (AI)-Generated Text} (see pp.\ 5--6 in the IEEE PSPB Operations Manual).

\bibliographystyle{IEEEtran}
\bibliography{over-desktop}

\begin{IEEEbiography}[{\includegraphics[width=1in,height=1.25in,clip,keepaspectratio]{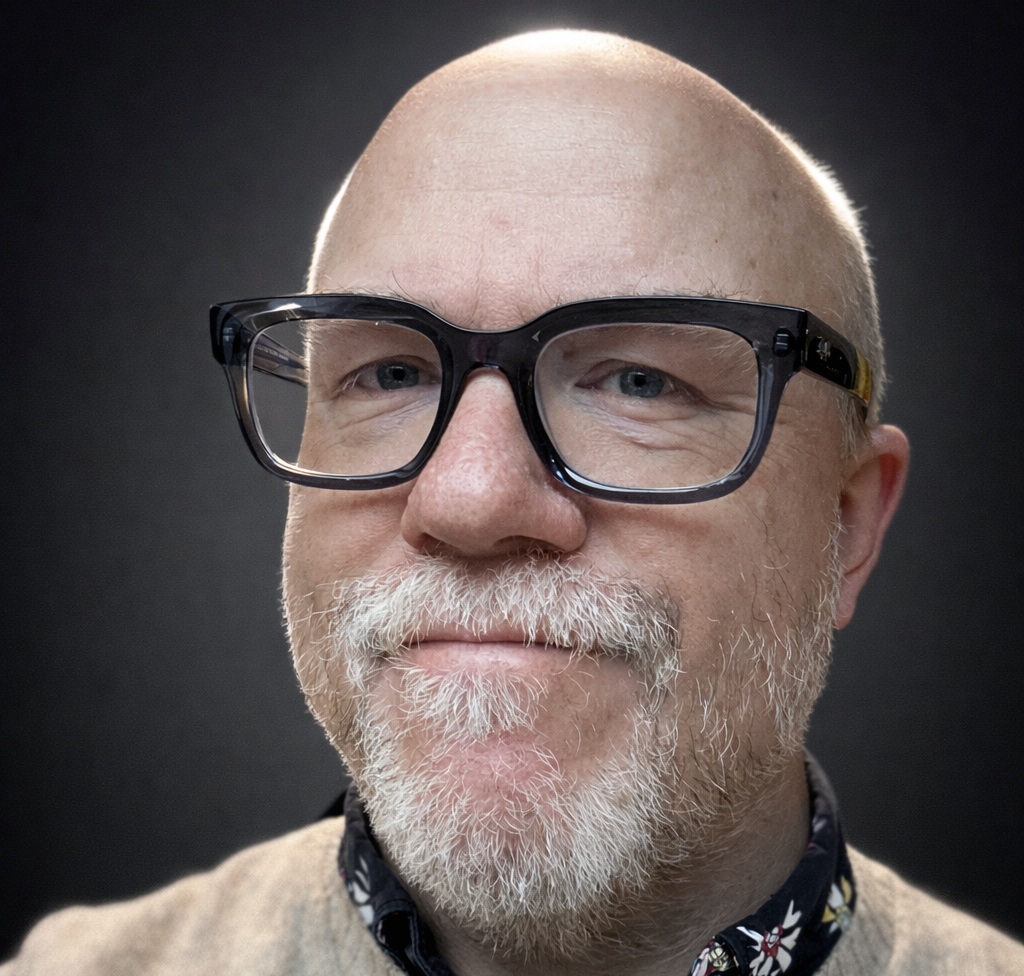}}]{Niklas Elmqvist}
received the Ph.D.\ degree in 2006 from Chalmers University of Technology in G\"{o}teborg, Sweden.
He is a Villum Investigator and professor in the Department of Computer Science at Aarhus University in Aarhus, Denmark.
He was previously faculty at University of Maryland, College Park from 2014 to 2023, and at Purdue University from 2008 to 2014. 
His research interests include visualization, HCI, and human-centered AI.
He is a Fellow of the IEEE and the ACM.
\end{IEEEbiography}

\begin{IEEEbiography}[{\includegraphics[width=1in,height=1in,clip,keepaspectratio]{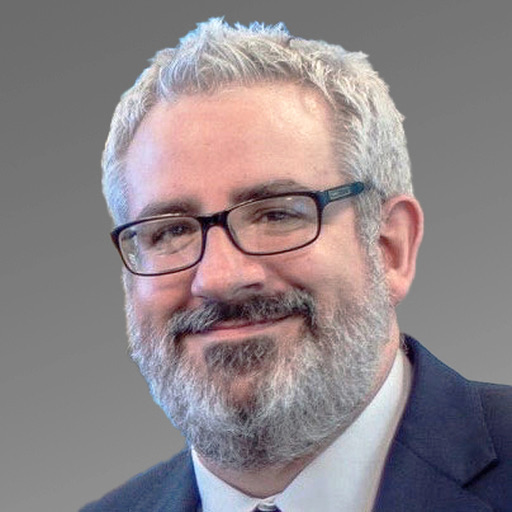}}]{Panagiotis D.\ Ritsos}
received the Ph.D. degree in 2006 from the University of Essex, Colchester, United Kingdom.
He is a Senior Lecturer in the School of Computer Science and Engineering, Bangor University, United Kingdom. 
His research interests include mixed and virtual reality, information visualization, and visual analytics. 
He is a Senior Member of the IEEE, the IEEE Computer Society, and a member of the ACM.
\end{IEEEbiography}

\begin{IEEEbiography}[{\includegraphics[width=1in,height=1in,clip,keepaspectratio]{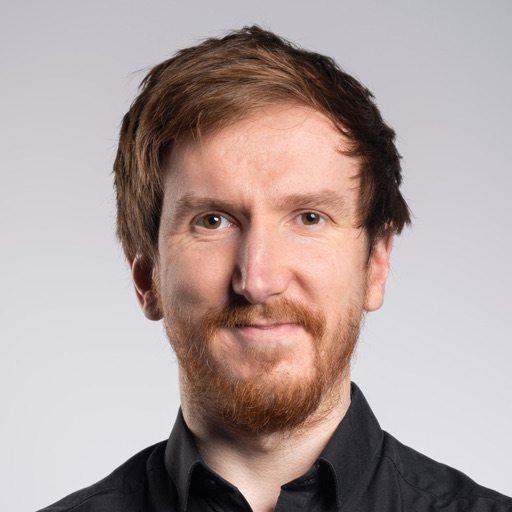}}]{Peter W.\ S.\ Butcher}
received the Ph.D. degree in 2020 from the University of Chester, UK.
He is a Lecturer in the School of Computer Science and Engineering, Bangor University, UK.
His research interests include human-computer interaction, information visualization, and immersive analytics.
\end{IEEEbiography}

\vfill

\end{document}